\documentclass[useAMS,usenatbib]{mn2e} 
\usepackage{epsfig}
\usepackage{graphicx,amsmath,multicol}
\usepackage{psfrag}
\usepackage[dvipsnames]{xcolor}
\definecolor{webgreen}{rgb}{0,.5,0}
\definecolor{webbrown}{rgb}{.6,0,0}
\usepackage[pdfpagelabels]{hyperref}
\usepackage[normalem]{ulem}
\hypersetup{%
   colorlinks=true,hyperfootnotes=false,%
   breaklinks=true,%
   plainpages=false, bookmarksnumbered, bookmarksopen=true,
   bookmarksopenlevel=1,%
   urlcolor=webbrown, linkcolor=RoyalBlue, citecolor=webgreen,%
}

\newcommand{\comments}[1]{}
\newcommand       \be           {\begin{equation}}
\newcommand       \ee           {\end{equation}}
\newcommand       \ba           {\begin{eqnarray}}
\newcommand       \ea           {\end{eqnarray}}
\newcommand       \grad         {\nabla}

\newcommand       \apj          {ApJ}

\newcommand       \apjs         {ApJS}

\newcommand       \mnras        {MNRAS}
\newcommand       \araa         {Ann. Rev. Astr. Astr.}

\def\msun{\rm \ M_\odot}

\def\lesssim{\mathrel{\hbox{\rlap{\hbox{\lower4pt\hbox{$\sim$}}}\hbox{$<$}}}}
\def\gtrsim{\mathrel{\hbox{\rlap{\hbox{\lower4pt\hbox{$\sim$}}}\hbox{$>$}}}}

\title[Turbulence \& cooling in cluster cores]
{Turbulence and cooling in galaxy cluster cores}  

\author[N.\ Banerjee, P.\ Sharma]
{Nilanjan Banerjee $^\dag$, Prateek Sharma$^\ddag$\\
$^\dag$ Department of Physics, Indian Institute of Technology, Kharagpur (nilanjan.banerjee@iitkgp.ac.in)\\
$^\ddag$Department of Physics and Joint Astronomy Program, Indian Institute of Science, Bangalore, India 560012 (prateek@physics.iisc.ernet.in)}

\begin{document}

\pagerange{\pageref{firstpage}--\pageref{lastpage}} \pubyear{2012}
\maketitle

\label{firstpage}

\begin{abstract}
We study the interplay between turbulent heating, mixing, and radiative cooling in an idealized model of cool cluster cores.  Active galactic nuclei (AGN) jets are expected to drive turbulence and heat  cluster cores. Cooling of the intracluster medium (ICM) and stirring by AGN jets are tightly coupled in a feedback loop. We impose the feedback loop by balancing radiative cooling with turbulent heating. In addition to heating the plasma, turbulence also mixes it, suppressing the formation of cold gas at small scales.  In this regard, the effect of turbulence is analogous to thermal conduction. For uniform plasma in thermal balance (turbulent heating balancing radiative cooling), cold gas condenses only if the cooling time is shorter than the mixing time. This condition requires the turbulent kinetic energy to be  $\gtrsim$ the plasma internal energy; such high velocities  in cool cores are ruled out by observations. The results with realistic magnetic fields and thermal conduction are qualitatively similar to the hydrodynamic simulations. Simulations where the runaway cooling of the cool core is prevented due to  {\em mixing} with the hot ICM show cold gas even with subsonic turbulence, consistent with observations. Thus, turbulent mixing is the likely mechanism via which AGN jets heat cluster cores. The thermal instability growth rates observed in simulations with turbulence are consistent with the local thermal instability interpretation of cold gas in cluster cores.
\end{abstract}


\section{Introduction}
Galaxy clusters are the largest virialized structures ($\sim 10^{14}-10^{15} \msun$) in the universe, consisting of hundreds of galaxies bound by the gravitational pull of dark matter. The intracluster medium (ICM) consists of plasma at the virial temperature, $10^7-10^8$ K.  Out of the total mass in galaxy clusters, only $\sim 15\%$ is baryonic matter, majority ($\gtrsim 80 \%$) of which is in the ICM and only a small fraction ($\lesssim 20 \%$) is in stars (e.g., \citealt{gon07}). The dark matter is responsible for setting up a quasi-static gravitational potential (except during major mergers) in the ICM, which along with cooling and heating decides the thermal and dynamic properties of the ICM (e.g., \citealt{piz05,mcn07,sha12,gas12}).

The  central number density in a typical ICM ranges from $0.1$  cm\textsuperscript{-3} in peaked clusters to $0.001$ cm\textsuperscript{-3} in non-peaked ones. The plasma in the ICM cools by radiative cooling so it is a strong source of X-rays with luminosity of about $10^{43}-10^{46}$ erg s\textsuperscript{-1}. The cooling time in dense central cores of some clusters is few $100$ Myr, much shorter than the cluster age ($\sim$ Hubble time). However, spectroscopic signatures of cooling (e.g., \citealt{tam01,pet03,pfab06}) and the expected cold gas and young stars are missing  (e.g., \citealt{edg01,odea08}). As a result, theoretical models and numerical simulations  without additional heating predict excessive cooling and star formation (e.g. \citealt{saro06,li12}). This is the well known cooling flow problem.

The simplest resolution for the lack of cooling is that the ICM is heated. This heating does not significantly increase the temperature of the ICM but instead roughly balances cooling losses in the core. Hence, the ICM is in rough {\em global} thermal equilibrium. Possible mechanisms for heating include mechanical energy injection from AGN jets and bubbles (see \citealt{mcn07} for a review),   turbulence in the ICM caused by galactic wakes (e.g., \citealt{kim05}), cosmic ray convection (e.g., \citealt{cha07}), and thermal conduction (e.g., \citealt{zak03}, and references therein). While non-feedback processes, e.g., conduction, can contribute to heating, we expect AGN feedback to become dominant in the cluster core (\citealt{guo08}). Moreover, non-feedback heating is globally unstable because of enhanced cooling at lower temperatures  (e.g., \citealt{sok03}). 
 
 The feedback heating mechanism may be outlined as follows. Since cooling occurs in  rough equilibrium with heating, only a fraction of the core cools (\citealt{pet03}) via the formation of multiphase gas if the cooling time of the hot gas is sufficiently short (e.g., \citealt{piz05,cav09,sha12}).  The cold multiphase gas increases the accretion rate onto the central black hole which powers a radio jet (e.g., \citealt{cav08}). Close to the black hole the jet is relativistic (e.g., \citealt{bir95,tom10}) but slows down as the low-inertia jet ploughs through the dense ICM core (e.g., \citealt{chu01,guo11}). The irregular jets generate weak shocks and sound waves (e.g., see \citealt{san07} for observations; \citealt{ste09} for simulations), thus heating the ICM. In addition, the buoyant bubbles can drive turbulence in the core (e.g., \citealt{cha07,sha09}), which can heat the cooling ICM by turbulent forcing and by mixing the outer hot gas with the inner cool core.

While numerical simulations of feedback jets have recently been successful in demonstrating thermal balance in cool cluster cores for cosmological timescales (e.g., \citealt{dub11,gas12}), there are several puzzles remaining to be answered. The biggest being, what mechanism heats the cool core? Is it turbulent heating, mixing of hotter ICM with the cooler core, or jet-driven weak shocks? This is a challenging question, so we focus on idealized models for the jet and the ICM in this paper. We posit that the anisotropic injection of mechanical energy by AGN jets is effectively (of course with an efficiency factor) converted into isotropic, small scale turbulence due to a dynamic ICM (e.g., \citealt{hei06,gas12}). Large vorticity and turbulence can be generated when jet-driven shocks interact with preexisting bubbles/cavities (\citealt{fri12}). Thus, in our models energy is deposited via homogeneous, isotropic  turbulence, with mechanical energy input. 

To prevent catastrophic cooling we balance radiative cooling with turbulent heating.  This heat input prevents catastrophic cooling of the core. However, local thermal instabilities can still result in the formation of localized multiphase gas, much like what has been observed (e.g. \citealt{fab08}). We investigate two classes of models: first, with uniform initial conditions similar to cool cores; second, with two regions with different densities/temperatures in pressure balance. The first model investigates if turbulent heating can balance radiative losses in the core and the second one focuses on turbulent mixing of gases at dissimilar temperatures (\citealt{kim03,voi04,den05} have considered phenomenological models of turbulent mixing in past). We require almost sonic velocities for turbulent heating to balance radiative cooling but our velocities are much smaller and consistent with observations if turbulent mixing is the primary heating mechanism of the cool core. While clearly very idealized, our models have minimal adjustable free parameters and are (astro)physically well-motivated. 

Although we focus on the application to cluster cores, there are obvious connections of our work to the interstellar medium (ISM), where turbulence provides significant pressure support (e.g., \citealt{bou90}). Most studies of the phases of the ISM focus on heating by photoelectric effect due to stellar and extragalactic UV photons (e.g., \citealt{wol03}). However, turbulent heating and mixing are expected to be important in the ISM (especially in regions shielded from photons); indeed recent numerical simulations try to incorporate the effects of turbulence on the phase structure of the warm/cold ISM (e.g., \citealt{vaz00,aud05}).

Our paper follows an approach very similar to \citet{sha10}, but with an important difference. While in \citet{sha10} the feedback energy was directly added to internal energy, here we add the energy in the momentum equation; i.e., work done by turbulent forcing goes into building up kinetic energy at all scales via a turbulent cascade, and is only  eventually converted into irreversible heating at small viscous scales. Such mechanical stirring more closely mimics the kinetic feedback of AGN jets in clusters. Very importantly, turbulent stirring also affects the nature of multiphase cooling; only length scales over which the turbulent mixing time is longer than the cooling time are able to condense out of the hot gas.

Earlier papers, \citet{mcc12,sha12}, studied models including gravity, with idealized thermal heating balancing cooling at every height. Turbulence generated in these simulations was solely due to buoyancy and cooling; there was no mechanical energy input mimicking jets/bubbles. It was found that mixing generated due to Kelvin-Helmholz instabilities at the interface of an overdense blob and the ICM prevents runaway cooling of the blob if the ratio of the local thermal instability timescale and the free-fall time ($t_{\rm TI}/t_{\rm ff}$) is smaller than a critical value. This conclusion was confirmed with realistic AGN jet simulations in \citet{gas12}.  Our present simulations allow us to take a closer look at multiphase cooling and stirring of cool cluster cores by AGN jets.

While it is possible to measure the radiative cooling time ($t_{\rm cool}$) from observations, we cannot measure the thermal instability timescale ($t_{\rm TI}$) directly from observations because it depends on the density dependence of the {\em microscopic} heating rate. Cluster observations show that extended cold gas is seen in clusters with $t_{\rm cool}/t_{\rm ff} \lesssim 10$ (see Fig. 11 in \citealt{mcc12}), which in the context of above models suggests that $t_{\rm TI} \approx t_{\rm cool}$; this is expected if the effective microscopic heating rate per unit volume is independent of density. We note that a simple density dependence of the microscopic heating rate does not capture the complexity of turbulent heating/mixing, but our numerical experiments are consistent with $t_{\rm TI} \simeq t_{\rm cool}$.  

Our paper is organized as follows. In section \ref{sec:numerics} we discuss our numerical setup and the associated equations. In section \ref{sec:results} we discuss the results from our simulations, and finally in section \ref{sec:conc} we discuss the astrophysical implications of our results.

\section{Setup}
\label{sec:numerics}

\subsection{Model equations}
We model the ICM plasma using the fluid approximation. We ignore self gravity because it has a negligible effect at high temperatures we are interested in ($\gtrsim 10^4$ K). The ICM is subjected to optically thin radiative cooling and turbulent heating by random velocity perturbations. The magnetohydrodynamic (MHD) equations governing this plasma are given by:

\begin{subequations}\label{eq:MHD}
\begin{align}
\frac{\partial\rho}{\partial t}+ \vec{\grad} \cdot (\rho \vec{v}) =0,\\
\frac{\partial(\rho\vec{v})}{\partial t}+ \vec{\grad} \cdot (\rho \vec{v} \otimes \vec{v}+P^{\ast}I - \vec{B} \otimes \vec{B} )= \vec{F},\\
\label{eq:total_energy}
\frac{\partial E}{\partial t}+ \vec{\grad} \cdot \{ (E + P^{\ast}) \vec{v} - (\vec{B} \cdot \vec{v}) \vec{B} + \vec{Q} \} = \vec{F}\cdot\vec{v} - \mathcal{L},\\
\frac{\partial \vec{B}}{\partial t} - \vec{\grad} \times ( \vec{v} \times \vec{B} )= 0,\\
P^{\ast} = P + \frac{\vec{B} \cdot \vec{B}}{2},\\
E = \frac{P}{\gamma - 1} + \frac{\rho \vec{v} \cdot \vec{v}}{2} + \frac{\vec{B} \cdot \vec{B}}{2},
\end{align}
\end{subequations}
where $t$, $\rho$, $\vec{v}$, $P$, $\vec{B}$, $E$, and $\gamma$ have their usual meanings. There are three non-ideal MHD terms ($\vec{F},~\mathcal{L},~\vec{Q}$): $\vec{F}$ represents turbulent forcing at a given location in the ICM, $\mathcal{L}~(\equiv n_e n_i \Lambda$, where $n_e/n_i$ are electron/ion number densities and $\Lambda(T)$ is the temperature-dependent cooling function) is the radiative loss term, and $\vec{Q}$ is the heat flux due to anisotropic thermal conduction given by
\be \label{eq:DIFF}
\vec{Q} = - \kappa _{\rm S} \hat{b} (\hat{b} \cdot \vec{\grad}) T ,
\ee
where $\hat{b} \equiv \vec{B} / \lvert \vec{B} \rvert$ is the magnetic field unit vector. The conductivity is chosen to be equal to the classical Spitzer conductivity (\citealt{spt65}),
\be \label{eq:CND}
\kappa _ {\rm S} = \frac{1.84 \times 10^{-5}}{\ln \lambda} T^{5/2} \rm{erg\ s ^ {-1} K ^ {- \frac{7}{2}} cm ^ {-1}}.
\ee
We use the standard value of $\ln \lambda = 37$. We carry out hydro and MHD runs, and some runs with anisotropic thermal conduction.
The list of runs with appropriate parameters is given in Table \ref{tab:tab1}.

\subsection{Numerical setup}

For a plasma in thermal equilibrium, which is not subject to gravitational stratification, thermally instability leads to condensation of cold gas from the hot phase. Nonlinearly, the  cold  gas collapses to an extremely small scale $\sim$ Field length (in the cold phase; \citealt{fld65,koy04,sha10}). Resolving such structures using realistic microscopic parameters is impossible as the separation of scales is enormous. Convergence of the cold phase is partly achieved if we truncate the cooling function at a small temperature floor $T_{\rm{cutoff}}$ (as also done by \citealt{mcc12}),
\be \label{eq:COOL}
\mathcal{L}= n_e n_i \Lambda(T) \mathcal{H} (T - T_{\rm{cutoff}}),
\ee
where $\mathcal{H}$ is the Heaviside function and we choose $T_{\rm{cutoff}}=10^6$ K for most runs (one of our runs, MA4, uses a cut-off at $10^4$ K to assess sensitivity to this parameter). This choice is  reasonable since runaway cooling happens once a blob cools below few$\times 10^6$ K. We use the cooling function $\Lambda(T)$ for an ionised plasma as given in Figure 1 of \citet{sha10}. To prevent negative temperatures, anisotropic thermal conduction in Eq. \ref{eq:MHD}c is implemented using limited averaging of transverse temperature gradients (\citealt{sha07}).

Turbulent forcing, $\vec{F}$ in Eq. (\ref{eq:MHD}b,c), is calculated as follows. At every timestep, statistically uniform, isotropic Gaussian random velocity perturbations in each direction $\delta v_i(\vec{k})$ are generated in $\vec{k}$ space, which have the following spectrum (from \citealt{sto98}),
\be \label{eq:SPEC}
| \delta v_i(\vec{k}) | \propto k^3 e^{- (4k/k_{\rm peak})},
\ee
where $k_{\rm peak}=2\pi/(10 {\rm kpc})$ is the fiducial peak driving wavenumber. Then, the velocity field is made divergenceless by projecting $\delta \vec{v}(\vec{k})$ perpendicular to $\vec{k}$. After converting the velocity perturbations to real space, a constant is subtracted from all three components of velocity perturbations such that no net momentum is added to the box, i.e., $\langle \rho  \delta \vec{v}(\vec{x} ) \rangle = \vec{0}$. Turbulent forcing,  given by 
\be
\label{eq:forcing}
\vec{F}(\vec{x},t)=\rho \delta \vec{v}(\vec{x},t)/\delta t, 
\ee
is scaled to maintain global thermal equilibrium, $\langle \vec{F}\cdot  (\vec{v} + \delta \vec{v}) \rangle = \langle \mathcal{L} \rangle$, where $\langle \rangle$ denotes volume averaging and $\delta t$ is the CFL timestep.  

We have modified the ATHENA MHD (\citealt{st08}) code to solve equations \ref{eq:MHD} through \ref{eq:forcing}. We use piecewise linear spatial reconstruction, the CTU/van-Leer integrator,  and periodic boundary conditions. We use the Roe solver for MHD runs and HLLC solver for hydrodynamic simulations. 

It is important to note that the turbulent force $\vec{F}$ appears twice in the MHD equations, once in the momentum  equation and then in the energy equation. These are implemented separately as split updates without modifying the main MHD integrator. Cooling and conduction are also applied in an operator split fashion; sub-cycling is implemented for the cooling and conduction substeps.

Most of our runs use a 3-D cartesian grid extending 40 kpc in each direction, with a resolution of $128^3$; some runs with 80 kpc box size and $256^3$ resolution are also carried out. Our uniform runs are  initialized with a temperature and density typical of a cool core: $T_0 = 0.78~{\rm keV},~n_e = 0.1~\rm{cm^{-3}}$; the cooling time for these parameters is $\approx 100$ Myr. In another set of runs (hereafter referred to as mixing runs) we use an 80 kpc domain with $(40 {\rm kpc})^3$ occupied by $n_e=0.1$ cm$^{-3}~(T=0.78$ keV) and rest of the volume by $n_e/3$ in pressure balance. Homogeneous, isotropic initial density perturbations are seeded with $( \frac{\delta \rho}{\langle \rho \rangle } ) _ {\rm{rms}} = 10^{-3}$ (the spectrum is the same as in \citealt{sha10}).  For MHD simulations a uniform magnetic field with $\beta= 100$ (ratio of plasma pressure and magnetic pressure) is initialized. The plasma composition is such that the mean mass per particle and electron are $~\mu = 0.62$ and $\mu _e=1.17$, respectively. 

\section{Results}
\label{sec:results}

We have carried out several simulations to study the interplay of radiative cooling, turbulent heating and mixing in detail. Simulations include hydro and MHD runs, with and without thermal conduction. Different runs are summarized in Table \ref{tab:tab1}. Various important results are discussed in the following sections.

\begin{table}
\caption{Various runs}
\resizebox{0.45 \textwidth}{!}{%
\begin{tabular}{ccccccc}
\hline
\hline
Label $^\dag$ & Res. & $L_x$ (kpc) & $\beta^\ddag$ & $\frac{\langle \rm{heating} \rangle}{\langle \rm{cooling} \rangle}$ & $t_{\rm TI}$ range (Myr) & $\alpha^{\dag\dag}$ range \\
 & & $=L_y=L_ z$ & & & & \\
\hline
H$^\ast$ & $128^3$ & 40 & ... & 1.0 & $59~{\rm to}~116$ & $-0.8~{\rm to}~0.6$ \\
Hh & $256^3$ & 40 & ... & 1.0 & $52~{\rm to}~132$ & $-1.2~{\rm to}~0.7$ \\
Ht & $128^3$ & 40 & ... & 1.0 & ... & ...  \\
Hs & $128^3$ & 20 & ... & 1.0 & ... & ...  \\
Hl & $128^3$ & 80 & ... & 1.0 & $100~{\rm to}~117$ & $0.3~{\rm to}~0.6$ \\
Hm & $128^3$ & 80 & ... & 1.0 & $66~{\rm to}~87$ & $-0.5~{\rm to}~0.1$  \\
Hha & $256^3$ & 40 & ... & no cooling & ... & ... \\
M & $128^3$ & 40 & 100 & 1.0 & 67~{\rm to}~69 & $-0.5~{\rm to}~-0.4$ \\
MA & $128^3$ & 40 & 100 & 1.0 & $73~{\rm to}~93$ & $-0.3~{\rm to}~0.2$ \\
MAm & $128^3$ & 80 & 100 & 1.0 & $66~{\rm to}~87$ & $-0.5~{\rm to}~0.1$\\
MA4 & $128^3$ & 40 & 100 & 1.0 & 31 to 52 & $-3.4~{\rm to}~-1.2$  \\
\hline
\end{tabular}} \\
{\textbf{Notes}} \\
{$^\dag$}{H stands for hydrodynamics, M is for MHD, h refers to high resolution, A represents anisotropic thermal conduction, and m stands for mixing runs. Run Ht has half (double) the fiducial density (temperature). Run Hs (Hl) uses half (double) the fiducial box size; $k_{\rm peak}$ (Eq. \ref{eq:SPEC}) is scaled with the box-size. Run Hha is an adiabatic high resolution run without cooling. The hydro mixing run Hm with the 80 kpc box-size has two regions: first a $(40~{\rm  kpc})^3$ zone at the fiducial temperature and density; rest of the volume is at thrice (a third) the fiducial temperature (density). The MHD mixing run MAm is similar. Run MA4 is a uniform MHD run with the stable phase at $10^4$ K; all other runs have stable phase at $10^6$ K.} \\ 
{$^\ddag$}{$\beta = \frac{8 \pi n k_B T}{B^2}$.}\\
{$^{\dag \dag}$ density dependence of heating using Eq. \ref{eq:TI} with $d \ln \Lambda/d\ln T=0$.}\\
{$^\ast$}{The fiducial run.}
\label{tab:tab1}
\end{table}

\subsection{The fiducial run: evolution \& convergence}
\begin{figure}
\begin{center}
\includegraphics[scale=0.48]{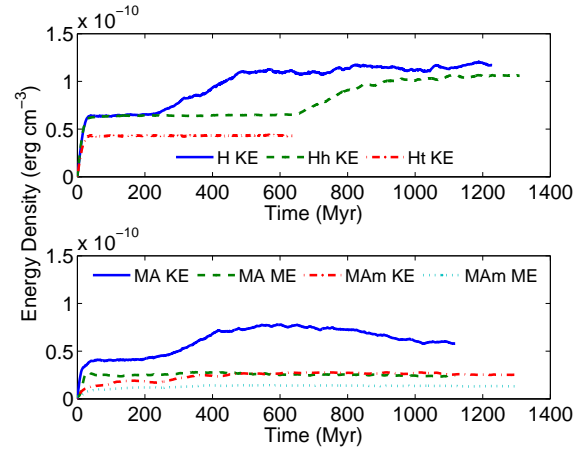}
\caption{{\em Top panel:} Volume averaged kinetic energy density for hydrodynamic runs: the fiducial run (H), the high resolution run (Hh), and the run with half the initial density (Ht). The lower density run does not show multiphase gas and the kinetic energy saturates without growth at later times, unlike other runs that show multiphase gas. {\em Bottom panel:} Kinetic and magnetic energy densities for the MHD runs with anisotropic thermal conduction, both the uniform run (MA) and the mixing run (MAm). Magnetic and kinetic energy densities for both MHD runs are smaller than in hydro runs because magnetic dissipation contributes to plasma heating.
}
\label{fig:ke}
\end{center}
\end{figure}

Most of our simulations, studying the interplay of radiative cooling and turbulent heating, show two stages in their evolution: first, where the impact of cooling is negligible and a turbulent steady state is attained; and second, where thermal instability takes over and leads to the formation of multiphase gas. The top panel of Figure \ref{fig:ke} shows the evolution of average kinetic energy density for the fiducial  hydro run (H) and its high resolution version (Hh). For these runs the kinetic energy first saturates at about $40$ Myr. This time corresponds to the eddy turnover time scale at the driving scale (corresponding to $k_{\rm peak}$ in Eq. \ref{eq:SPEC}) when a turbulent steady state is reached. At later times, around $300-600$ Myr, the ICM enters a second non linear stage of evolution, with a larger magnitude of kinetic energy. This increase is associated with the production of multiphase gas due to local thermal instability. Since radiative cooling is more efficient at lower temperatures and we impose thermal balance, larger kinetic energy needs to be dissipated to balance cooling losses.

\begin{figure*}
\centering
\includegraphics[scale=0.75]{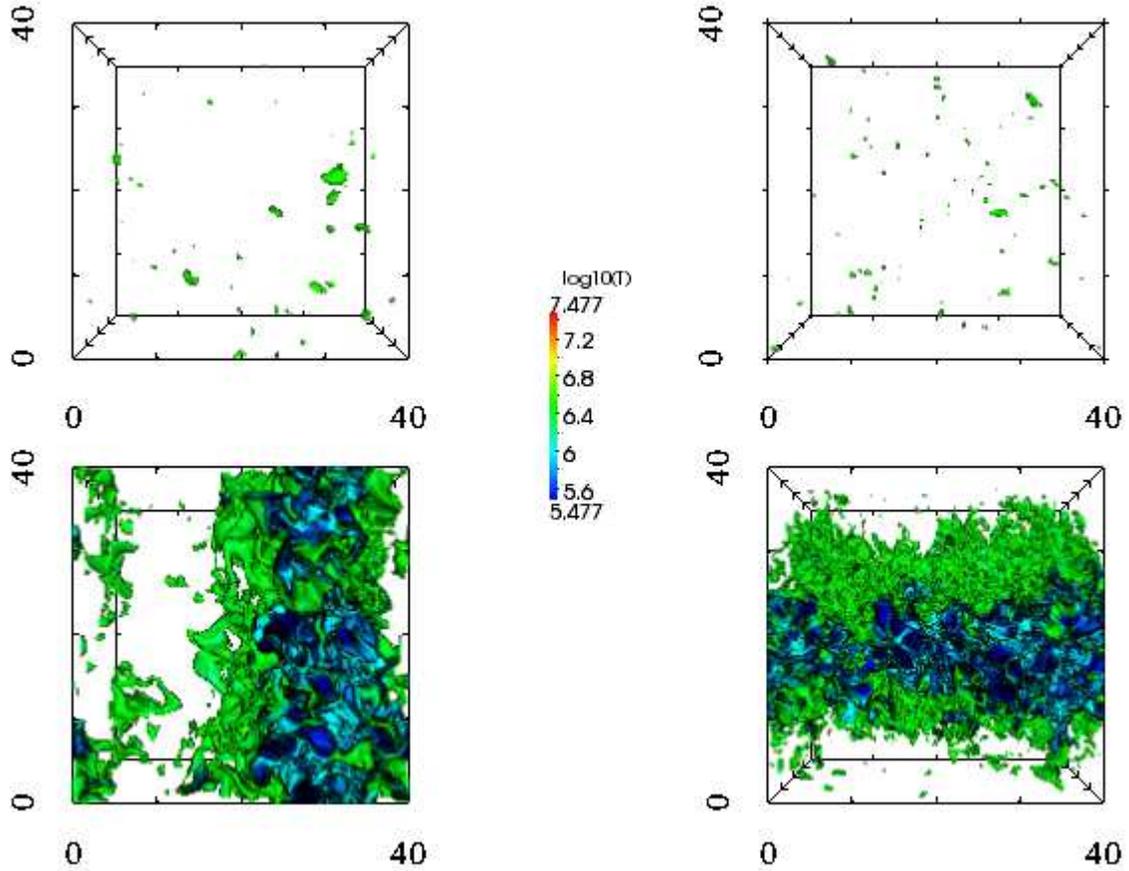}
\caption{Volume rendering of temperature ($T$) at $90$ Myr (top) and at $1.05$ Gyr (bottom) for the fiducial run ($128^3$; H; left) and $256^3$ (Hh; right) run. The box size is 40 kpc. The colorbar shows  the temperature; all the plasma above $3\times 10^6$ K is transparent and the plasma below this temperature is opaque. The first nonlinear state (at 90 Myr) shows  perturbations due to driven turbulence; the second nonlinear state at 1.05 Gyr is dominated by cold gas. The temperature distributions for the two resolutions are statistically similar. There is substantial gas at intermediate temperatures that forms a turbulent sheath around the coolest clumps.
\label{fig:snap}}
\end{figure*}

\begin{figure*}
\centering
\psfrag{a}[cc]{{\large $\frac{dM}{d\log_{10}T} (\msun)$}}
\psfrag{b}[cc]{{\large $\frac{dM}{d\log_{10}{\cal M}} (\msun)$}}
\includegraphics[scale=0.6]{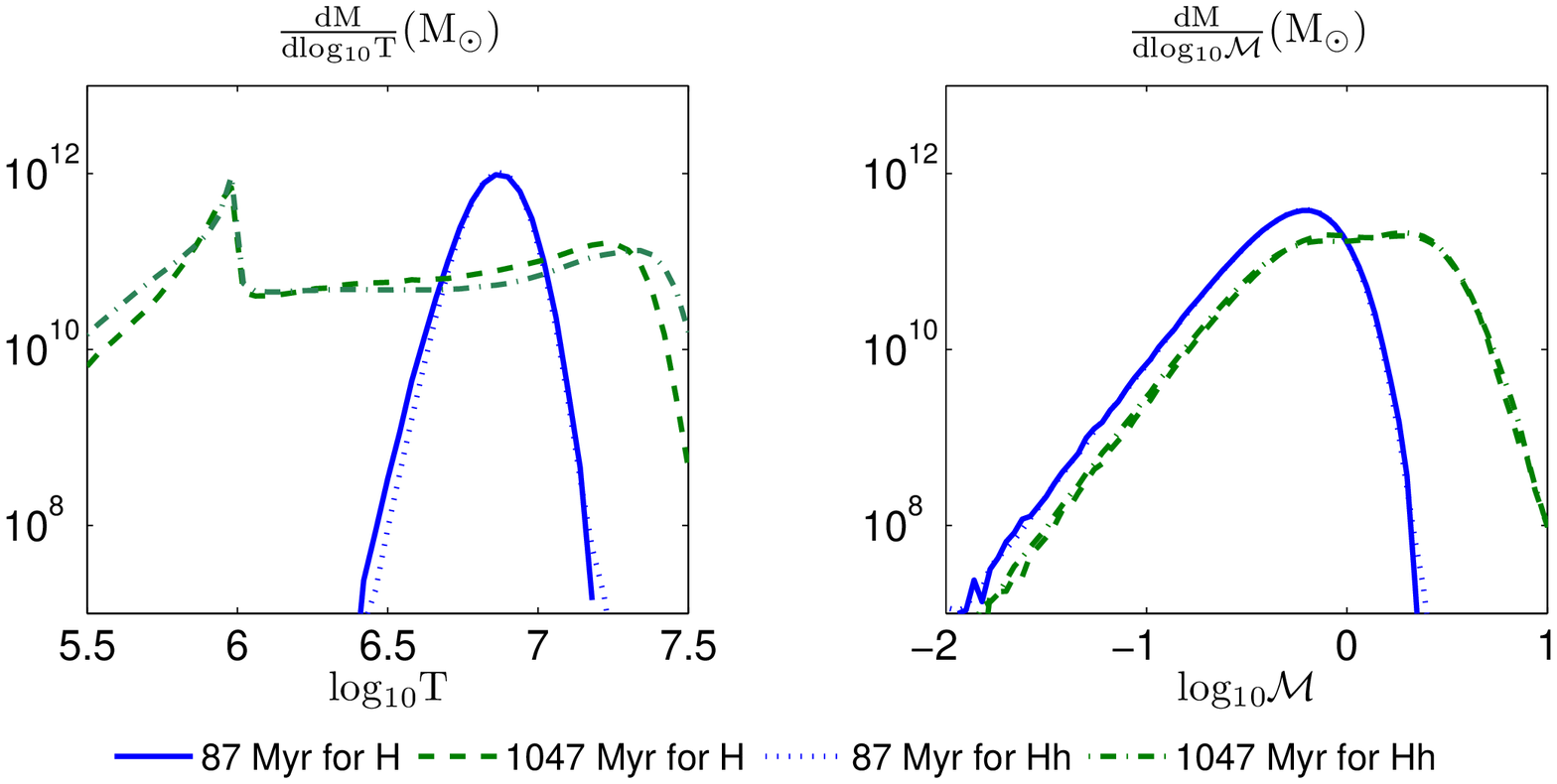}
\caption{The probability distribution function (pdf) of mass as a function of temperature and Mach number (${\cal M} \equiv v/[\gamma P/\rho]^{1/2}$) at two different times for the fiducial run (H) and its high resolution version (Hh). There is a clear excess of gas at the temperature of the thermally stable phase.}
\label{fig:pdf_fid}
\end{figure*}

Figure \ref{fig:snap} shows the volume snapshots of temperature at $90$ Myr (saturated turbulence without cold gas) and at $1.05$ Gyr (non linear stage with multiphase gas). Snapshots from the fiducial run are compared with a similar higher resolution ($256^3$) run. It is obvious from these plots that significant amount of cold gas has condensed by $1.05$ Gyr. The snapshots for different resolutions look different because the random number generating seed is the same but the number of grid points is 8 times larger for the higher resolution run. The features show a much larger range in the higher resolution run because of a smaller dissipation scale. 

Statistically the simulations at two resolutions are similar, as can be seen from the probability distributions of temperature and Mach number in Figure \ref{fig:pdf_fid}. The cooler gas starts as small clouds which are mixed and merged, and at later times only large clouds comparable to the box size survive. The coolest, densest gas is clumpy and covered by a sheath of lower density, higher temperature gas. These clouds are morphologically distinct from the filaments observed in cool core clusters (e.g., \citealt{fab08}). This is because the cold filaments are magnetically dominant and not prone to mixing, unlike our clouds in Figure \ref{fig:snap}. Magnetic fields are expected to be dominant in the cold phase because of flux freezing in dense filaments that are much denser (by a factor $\sim10^3$) than the ambient ICM. Moreover, turbulence is not expected to be as vigorous as in our hydro runs if other sources of heating such as thermal conduction and turbulent mixing (with the hotter gas) are present. We discuss the formation of cold filaments further in section 4.3.

The left panel of  Figure \ref{fig:pdf_fid} shows the temperature distribution of mass, which peaks at $10^6$ K (the thermally stable phase) and at $\sim 10^7$ K, but there is substantial mass at intermediate temperatures; this is also seen in Figure \ref{fig:snap}.  The mass at the thermally stable phase slowly build up and eventually saturates. The right panel of Figure \ref{fig:pdf_fid} shows the Mach number distribution at 90 Myr and 1.05 Gyr. Unlike the temperature distribution, the Mach number distribution shows a single broad peak. The Mach number peak at ${\cal M} \gtrsim 1$ disagrees with the observations of cool core clusters that show subsonic gas. A comparison of temperature and Mach number distributions for the two resolutions show that the results are statistically converged. Since the results show statistical convergence at $128^3$ and because the simulations with cooling are expensive, most of our other runs with magnetic fields and conduction are carried out at the resolution of $128^3$. 

\begin{figure}
\centering
\psfrag{a}{{\large $\left ( \frac{\delta \rho}{\rho} \right)_{\rm rms}$}}
\includegraphics[scale=0.48]{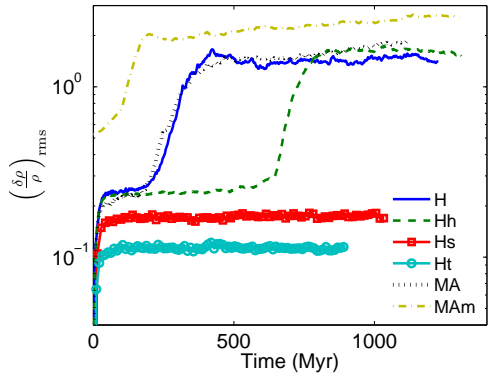}
\caption{The volume-averaged rms density fluctuations ($[\delta \rho/\rho]_{\rm rms}$) as a function of time for different runs. The figure clearly shows two stages of evolution: first dominated by turbulence and the second by thermal instability. The linear growth rates measured using these plots (where rms density fluctuations rise from an amplitude of 0.3 to 1) are given in Table \ref{tab:tab1}. The runs that do not show cold gas saturate with lower density perturbations. 
\label{fig:rho}}
\end{figure}

The amplitude of density perturbations is a good diagnostic to characterize the two nonlinear stages in the evolution of our simulations. In the turbulent stage, the density fluctuations are small because turbulence is subsonic. However, when local thermal instability kicks in later, the density perturbations become highly nonlinear because of condensation of cold gas. Figure \ref{fig:rho} shows the volume-averaged rms density fluctuations ($[\delta \rho/\rho]_{\rm rms} = \sqrt {\langle [\rho -\langle \rho \rangle]^2 \rangle}/\langle \rho \rangle $, where $\langle \rangle$ denotes volume averaging) as a function of time for the fiducial run, and for other selected runs. It is clear that $(\delta \rho/\rho)_{\rm rms}$ steadily increases to reach a saturation between $ \approx 0.2-0.3$ at $t \approx 40$ Myr, which is approximately the eddy turnover time. After some time in this nonlinear turbulent steady state with small density perturbations, the density perturbations start growing roughly exponentially and saturate with $(\delta \rho/\rho)_{\rm rms} \gtrsim 1$ at later times. 

\begin{figure*}
\centering
\includegraphics[scale=0.5]{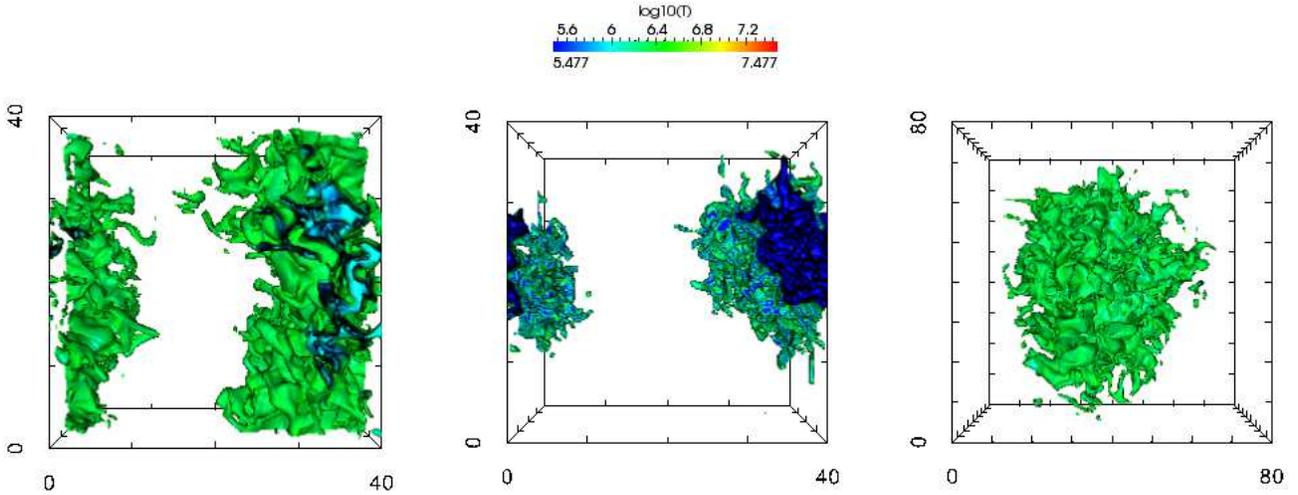}
\caption{Volume rendering of temperature ($T$) at $1.05$ Gyr for the MHD runs  MA, MA4 (run with the stable phase at $10^4$ K), and MAm (mixing run; from left to right; see Table \ref{tab:tab1}). The colorbar shows the temperature; all the plasma above $3\times 10^6$ K is transparent and the plasma below this temperature is opaque. The box-size is 40 kpc for MA and MA4, but 80 kpc for the mixing run MAm. Overall, the volume rendering plots look `cloudy,' similar to the hydro runs (Fig. \ref{fig:snap}) but there are some magnetically-dominated filamentary structures at cloud boundaries.
\label{fig:snap_MHD}}
\end{figure*}

The growth rate of density perturbations due to thermal instability can be measured from Figure \ref{fig:rho}. The measured thermal instability growth rate can help us compare our idealized simulations with cluster observations, as we discuss in section 4.2. A very precise measurement of the thermal instability growth rate is not possible from our simulations because the initial saturation of $\frac{\delta \rho}{\langle \rho \rangle}$ (due to turbulence before thermal instability takes over) happens at a large value. The range of measured thermal instability growth time scales for different runs are given in Table \ref{tab:tab1}.

\subsection{Impact of driving scale, magnetic fields and conduction}
 
 After discussing the results from hydro runs in the previous section, we study the influence of numerical parameters and additional physics such as MHD and thermal conduction on our results.
 The run with half the box-size (20 kpc on a side; run Hs in Table \ref{tab:tab1}) and a similarly scaled driving scale does not show cold multiphase gas but the run with double the box-size (and a larger turbulence driving scale; run Hl in Table \ref{tab:tab1}) shows condensation of multiphase clouds. The absence of multiphase gas for the run with a smaller driving scale is understandable. For the smaller driving scale run which is pumping in similar energy per unit volume as the run with a larger driving scale (energy input per unit volume is similar because radiative cooling per unit volume is fixed), the velocity at the driving scale is smaller ($\propto L^{1/3}$, where $L$ is the driving scale) and the mixing time ($\propto L^{2/3}$) at the driving scale is shorter. In this argument we assume that the density perturbations at scales larger than the driving scale are negligible. The scales smaller than the driving scale are mixed at an even faster rate but the cooling rate is independent of length scale. Thus, the runs with smaller driving scales are less susceptible to show multiphase gas. Since the small box run does not show multiphase gas, it also does not show the late time increase in kinetic energy (Fig. \ref{fig:ke}) and $(\delta \rho/\rho)_{\rm rms}$ (Fig. \ref{fig:rho}). The lower density uniform run with a long cooling time (Ht in Table \ref{tab:tab1}) also does not produce multiphase gas because cooling time is longer than the mixing time; this run behaves like the smaller box run Hs.

The runs with magnetic fields in the lower panel of  Figure \ref{fig:ke} show that the kinetic energy for MHD runs is smaller compared with the pure hydro runs; this is true for both runs with and without conduction, although only runs with conduction are shown. The kinetic energy density is lower because the gas is not only heated due to the dissipation of kinetic energy but also due to the dissipation of turbulent magnetic energy, and hence the energy requirement on turbulence is smaller. Since turbulent velocity is smaller with magnetic fields and the mixing time is longer, cold gas is expected to condense more easily in MHD runs. Moreover, magnetic fields provide resistance to mixing of cooler gas due to turbulence. The left panel in Figure \ref{fig:snap_MHD} shows the temperature snapshot at late times for the uniform MHD run with anisotropic conduction (MA). Globally the features look very similar to the hydro run but the cooler gas is a bit more filamentary because of magnetic fields. Small scale features are also suppressed because of thermal conduction. The Field length is $\approx 10$ kpc but conduction is suppressed perpendicular to magnetic field lines, and nonlinearly conduction does not seem to play an important role in the formation of multiphase gas.

Figure \ref{fig:pdf} shows the temperature pdf of gas for the fiducial hydro run and the MHD runs at early and late times. The pdfs at early times for the uniform runs are very similar. At later times the fraction in  cold gas is larger for the uniform MHD run as compared to the hydro run. Since energy injection rate equals average cooling rate in our set up, the temperature of the hot phase is higher for MHD because of enhanced radiative cooling and a larger cold fraction. The fraction of gas at intermediate temperatures is also smaller in MHD because of weaker turbulence. Figure \ref{fig:pdf_mix} shows the Mach number pdfs of gas for hydro and MHD runs. Unlike the fiducial hydro run, which shows a broad Mach number distribution with a single peak (although on a closer look one can see two peaks even for the hydro run), the uniform MHD run with conduction shows two clear peaks in the Mach number distribution. Both the subsonic and supersonic peaks have a lower Mach number compared to their hydro counterparts, which is consistent with a smaller kinetic energy for MHD runs (Fig. \ref{fig:ke}). The higher Mach number peak corresponds to the cooler thermally stable gas and the lower Mach number peak corresponds to the hot peak in the temperature distribution. As expected from Figure \ref{fig:pdf}, the fractional mass in the higher Mach number peak is larger for MHD as compared to the hydro run. Figure \ref{fig:rho} shows that the rms density perturbations in MHD runs behave like the hydro runs, and therefore the thermal instability growth rate estimates in Table \ref{tab:tab1} are also very similar.

\subsection{Mixing runs}

Since our uniform runs give large velocities in the hot gas (Figs. \ref{fig:pdf_fid}, \ref{fig:pdf_mix}), they are inconsistent with observations of cool-core clusters which show subsonic velocities. The Mach numbers are a bit lower for MHD runs, but still too big compared to observations. In order to make our setup consistent with observations we carry out mixing runs (Hm and MAm in Table \ref{tab:tab1}) where a fraction of the gas is at the temperature corresponding to the cool core and rest of the volume is occupied by the hotter (by a factor of 3; this choice is motivated by the fact that most cluster cores show dramatic lack of gas below 3 times the peak temperature; \citealt{pet03}), dilute ICM. In this setup heating of the cool core happens mainly through turbulent mixing of the hotter and cooler gases, but not via turbulent heating. Unlike uniform runs, there is transport (both conductive and turbulent) of heat from hotter ICM to the cooler gas. This setup mimics the mixing of hot ICM with the cool core driven by AGN jets.\footnote{ICM core can also be heated by the mixing of cosmic rays within the bubble with the cool core but cosmic ray particles are collisionless, so the interaction between cosmic rays and thermal plasma has to be mediated via magnetic fields (e.g., \citealt{guo08}). Turbulent mixing should be similar for the mixing of the cool core with either cosmic rays or the hotter ICM.} Again, global thermal balance is imposed. Since most volume is occupied by the hot/dilute plasma, the turbulent energy input is much smaller compared to uniform runs, and is in line with observations.

Lower panel of Figure \ref{fig:ke} shows that the kinetic energy density in the MHD mixing run (MAm) is about a factor of 5 smaller than in the fiducial hydro run. Low kinetic energy suppresses mixing and makes it easier for cold gas to condense. Figure \ref{fig:pdf_mix} shows the volume and mass Mach number distribution for the mixing run and other uniform runs. In the nonlinear state, there are two peaks in the Mach number distribution. The lower Mach number peak at ${\cal M}\approx 0.2$ corresponds to the subsonic hot phase and the ${\cal M}>1$ peak corresponds to the cold phase. Thus the hot phase Mach number is consistent with observations of cool core clusters; cold gas is expected to be supersonic.

The rms density amplitude in Figure \ref{fig:rho} for the mixing run increases earlier compared to the other runs because the cooler region cools on average, since the cooling rate of the cooler region is the same as our fiducial run but the heating rate is smaller by $\approx$0.3. The growth timescale is similar to the uniform runs (see Table \ref{tab:tab1}). The right panel of Figure \ref{fig:snap_MHD} shows the volume rendered temperature plot for the mixing run. Note that the box size is double the fiducial value and the volume occupied by the cooler gas is larger. Moreover, the coolest gas is completely covered by a sheath of warmer gas. A slight excess of cold gas in mixing runs is also seen for temperature pdfs in Figure \ref{fig:pdf_mix}. The amount of cold gas in all runs with multiphase gas is $\sim 10^{10} \msun$, comparable to observations. This is mainly because of our choice of parameters; a quantitative understanding of mass in cold phase involves interplay of local thermal instability and  gravity (e.g., \citealt{sha12}).

\subsection{Multiphase gas}
\begin{figure*}
\centering
\includegraphics[scale=0.6]{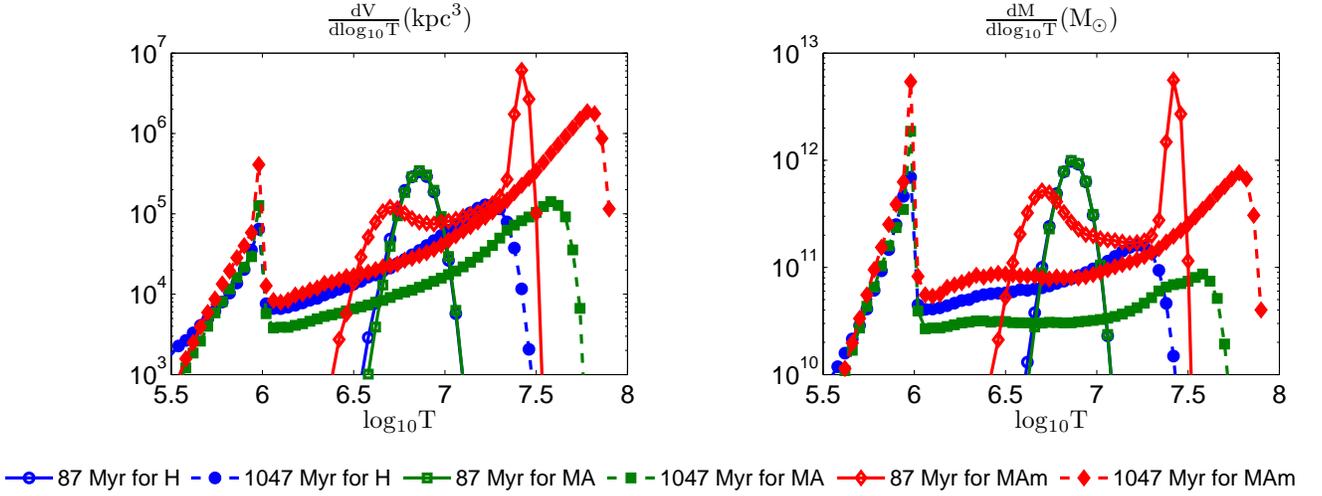}
\caption{Probability distribution functions of  mass ($\frac{\rm{dM}}{\rm{d} \log _{10} \rm{T}}$) and volume ($\frac{\rm{dV}}{\rm{d} \log _{10} \rm{T}}$) with respect to temperature at early and late times for the uniform  hydro run (H), the uniform MHD run with anisotropic conduction (MA), and the MHD mixing run with anisotropic conduction (MAm). The temperature distribution is bimodal after a thermal instability timescale; the bimodality is sharper for MHD runs with conduction. Note that the pdfs at early times for runs H and MA are almost coincident.
\label{fig:pdf}}
\end{figure*}

Figure \ref{fig:pdf} shows the probability distribution of mass and volume as a function of gas temperature in the linear and nonlinear stages of the development of thermal instability for different runs. Initially all the gas is hot with small temperature dispersion. After a cooling time a second peak develops at the thermally stable temperature. In the nonlinear stage of thermal instability for the fiducial run (H) the colder peak becomes smoother and there is significant mass and volume occupied by gas at intermediate temperatures. The runs with magnetic field and conduction (MA and MAm) have sharper peaks at the thermally stable temperature because magnetic fields and anisotropic conduction shields cold filaments from disruption. Clearly the mass and volume fraction in the hot phase is larger in all cases. This means that none of these runs are undergoing global cooling catastrophe even after several cooling times.

For the mixing run (MAm) in Figure \ref{fig:pdf} at early times there is a peak at $\sim 1$ keV corresponding to the initial cool core. In the nonlinear stage, this thermally unstable plasma either cools to $10^6$ K or becomes a part of the hot phase. The hot phase temperature increases because the hot zone is heated on average. The pdfs corresponding to the mixing runs are qualitatively similar to the observations of cool core clusters. Although not shown in Figure \ref{fig:pdf}, our run with the thermally stable phase at $10^4$ K shows the low temperature peak at this temperature. Thus, for a realistic cooling curve we expect large abundance of gas at the temperature below which the cooling function decreases abruptly. This gas at thermally stable temperature corresponds to H$\alpha$ emitting filaments observed in cool core clusters. The middle panel of Figure \ref{fig:snap_MHD} shows the volume rendered plot of temperature for the uniform MHD run with a stable phase at $10^4$ K (run MA4). The volume occupied by the coolest and intermediate temperature gas is smaller compared to the fiducial hydro and MHD runs because the gas is cooler and hence occupies smaller volume. The cold gas is still cloudy and not filamentary. The thermal instability growth timescale for the MA4 run in Table \ref{tab:tab1} shows a much smaller value compared to the other runs because the cooling time becomes quite short for smaller temperatures. In MA4 runs the gas cools rapidly through $\sim 10^5$ K, the temperature at which the cooling function peaks.

\section{Interpretation \& conclusions}
\label{sec:conc}

In this paper we focus on turbulent heating/mixing as a mechanism via which the mechanical energy is thermalized, using an idealized well-posed setup adhering to the phenomenological model where  cooling in the core is roughly balanced by average energy injected through turbulence. The model assumes uniform, isotropic distribution of turbulence, and global thermal equilibrium in the ICM core. While non-turbulent mechanisms, such as  thermal conduction (thermal conduction is expected to be suppressed substantially because of the wrapping of magnetic fields perpendicular to the radial direction; e.g., \citealt{par09,wag13}) and cosmic ray streaming (e.g., \citealt{guo08a}), can heat the cluster core, AGN jet driven turbulence is expected to be the dominant heating mechanism.

In reality, the interaction of the AGN jet with the ICM is expected to be rather complicated but small scale heating should be qualitatively similar to our idealized model. At large scales buoyancy forces, which are independent of the scale, are important but as we go to small scales turbulent forcing becomes more important; the effect of global stable stratification is even more easily overcome for a thermally conducting plasma  such as the ICM (e.g., Fig. 11 in \citealt{sha09}). 

In previous idealized models (\citealt{sha10,sha12}) we added heating as a term in the thermal energy equation. This is very idealized because in reality heating involves turbulent motions in a fundamental way. Turbulence can stir up the ICM and suppresses the formation of cold gas, especially at small scales (e.g., \citealt{gas13}). In idealized setups turbulence is weaker and is {\em generated by} thermal instability in presence of gravity, but in reality there is turbulence stirred by AGN jets which heats and mixes the ICM. 

\subsection{Cold gas condensation with turbulent heating \& mixing}

The formation of cold gas in uniform gas is determined by the ratio of the cooling time $t_{\rm cool}~(\equiv 1.5nk_BT/n_en_i\Lambda$ ; assuming that the thermal instability timescale $t_{\rm TI} \approx t_{\rm cool}$) and the mixing time $t_{\rm{mix}}$.  The ratio $t_{\rm cool}/t_{\rm mix}$ is a scale dependent quantity, which increases with a decreasing length scale because $t_{\rm mix} \equiv l/v_l$ is shorter at smaller scales ($l$ is the length scale and $v_l$ is the velocity at this scale). If turbulent heating balances cooling, then
\be
\label{eq:scaling_MP}
\dot{E}_{\rm turb} \approx \rho v_L^3/L = \rho v_L^2/ t_{{\rm mix},L} \approx n_i n_e \Lambda = U/t_{\rm cool},
\ee
where $L$ is the driving scale and $U=P/(\gamma -1)$ is the internal energy density. This energy balance equation implies that
\be
\label{eq:cond_MP}
t_{\rm cool}/t_{{\rm mix},L} \approx U/2K,
\ee
where $K \equiv \rho v_L^2/2$ is the kinetic energy density at the driving scale. 

According to Kolmogorov scaling in subsonic turbulence, for scales smaller than the driving scale, $v_l \propto l^{1/3}$ and $K \propto l^{2/3}$; thus $t_{\rm cool}/t_{\rm mix}$ decreases with an increasing length scale. The scales larger than the driving scale ($l>L$) have negligible velocities;  transport at these scales happens due to eddies of size $L$.\footnote{We are grateful to the anonymous referee for drawing our attention to scales larger than $L$, and their importance for generating multiphase gas. Thus the mixing time at $l \gtrsim L$ is given by $t_{{\rm mix}, l>L} \approx l^2/(L v_L) = (l/L)^2 t_{{\rm mix},L}$, which can be significantly longer than the mixing time at the driving scale.}

 Multiphase gas can condense out only at scales where $t_{\rm cool}/t_{\rm mix} \lesssim 1$ (otherwise cooling blobs are mixed before they can cool to the stable temperature). 
 Cold gas can condense out most easily at scales larger than the driving scale for  $t_{\rm cool} \lesssim t_{{\rm mix}, l>L}$, or equivalently, for  $2K/U \gtrsim (L/l)^2$.
 In case the driving scale ($L$) is comparable to the size of the cool core, cold gas condenses only if $2K/U \approx {\cal M}^2 \gtrsim 1$; i.e., if the driving velocity is $\gtrsim$ the sound speed. The size of AGN jets is typically $\gtrsim$ the cluster core, and therefore ${\cal M}^2 \gtrsim 1$ is required for cold
 gas condensation in a uniformly stirred core. The Mach number pdf at late times in Figure \ref{fig:pdf_fid} indeed shows a broad peak at Mach number of unity. Because of large turbulent motions, the cold gas in uniform runs comprise of large clouds, and not slender filaments as observed in cluster cores. Observations of cool core clusters show that the Mach number in cool core clusters (at temperatures traced by diagnostic lines) is $\lesssim 0.4$ (\citealt{wer09,san10}). Therefore, heating of the cool core due to turbulent dissipation (with driving at the scale length of cluster cores) is ruled out. However, mixing of hotter and cooler gas, driven by AGN jet turbulence and thermal conduction, can still heat the cool core without large turbulent velocities, as we discuss later. For lower mass halos, such as groups and individual galaxies, the ``core" size is much bigger (e.g., \citealt{sha12b}) than the stirring scale (due to supernovae and AGN) and cold gas can condense out for ${\cal M} < 1$.

Turbulent mixing, like thermal conduction, suppresses thermal instability at small scales. Consider a uniform medium where turbulent heating (assuming stirring at largest scales) balances cooling globally.  With conduction, the Field length is the length at which thermal diffusion timescale equals the thermal instability timescale ($t_{\rm TI}$; Eq. 8 in \citealt{sha10}). The turbulent Field length should be estimated by equating the turbulent mixing time $l/v_l = l^{2/3}L^{1/3}/v_L $ (here we have used Kolmogorov scaling; $v_l^3/l =$ constant, irrespective of scale) and $t_{\rm TI}$; i.e., $l_{F, {\rm turb}} = L^{-1/2} (v_L t_{\rm TI})^{3/2} \approx c_s t_{\rm cool}$ (assuming global thermal balance). Thus, only scales with ${\cal M} \gtrsim 1$ are thermally unstable. If stirring is at scales smaller than the box-size, the mixing time is longer by a factor $(l/L)^2$ for $l>L$, and $l_{F, {\rm turb}} \approx L (t_{\rm cool}/t_{{\rm mix},L})^{1/2} \approx L (U/2K)^{1/2} > L$. 


\begin{figure*}
\centering
\includegraphics[scale=0.6]{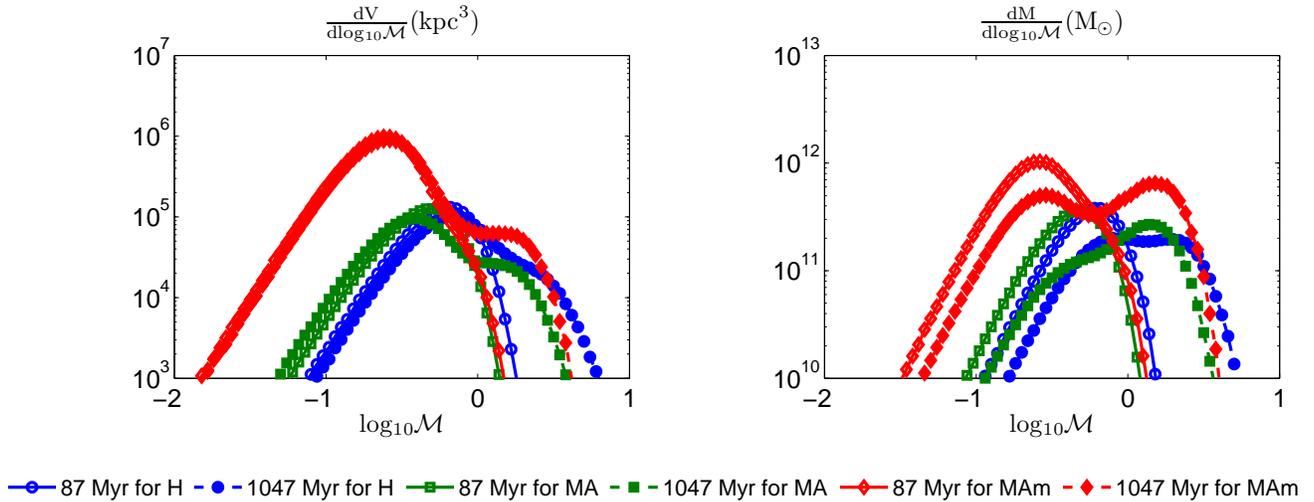}
\caption{Probability distribution functions of  mass ($\frac{\rm{dM}}{\rm{d} \log _{10} \rm{T}}$) and volume ($\frac{\rm{dV}}{\rm{d} \log _{10} \rm{T}}$) with respect to temperature at early and late times for the uniform  hydro run (H), the uniform MHD runs with anisotropic conduction (MA), and the mixing MHD run with anisotropic conduction (MAm). The temperature distribution is bimodal after a thermal instability timescale; the bimodality is sharper for MHD runs with conduction.
\label{fig:pdf_mix}}
\end{figure*}

To understand the mixing runs consider a setup with two zones in pressure equilibrium with temperature $T_0$ (density $n_0$) and $fT_0$ ($n_0/f$); the cooler zone occupies a volume fraction $f_v$. For our mixing runs $f=3$ and $f_v=1/8$. Now we will estimate the turbulent velocities required to balance cooling in the cooler zone. The turbulent energy injection rate per unit volume $\rho v_L^3/L$ (which is equal in hotter and cooler regions) for global thermal balance is $g n_0^2 \Lambda_0$, where $g=\{ f_v + (1-f_v)f^{-3/2}\} $ and we have assumed $\Lambda \propto T^{1/2}$. The net cooling rate of the cooler zone is, therefore, $n_0^2 \Lambda_0[1- g ]$. Now we want to estimate the rate at which turbulent mixing can bring heat from the hotter to the cooler regions. The turbulent velocity in the hot zone is obtained by noting that ${\rho_0 v_{L,{\rm hot}}^3}/(f L) = g n_0^2 \Lambda_0 $. This gives the turbulent velocity on the driving scale in the hot zone $v_{L,{\rm hot}} \approx (fg)^{1/3} c_{s0}^{2/3} (L/t_{\rm cool,0})^{1/3}$, where $c_{s0}$ is the sound speed in the cooler zone and $t_{\rm cool,0}$ its cooling time. Since the hot zone is overheated and the cooler zone is cooling on average, there is a flow of energy from the hotter to cooler zone and flow of mass in the opposite direction. 

The total energy equation, Eq. \ref{eq:total_energy}, in absence of magnetic fields and thermal conduction, can be simplified to
\be
\label{eq:e_simple}
\frac{\partial E}{\partial t}+ \vec{\grad} \cdot \{ (E + P) \vec{v} \} = \vec{F}\cdot\vec{v} - \mathcal{L},
\ee
where the first term on the right hand size represents turbulent heating (work done by turbulent force that is dissipated as heat in steady state) and the second term on the left hand size represents heating due to turbulent mixing. Integrating Eq. \ref{eq:e_simple}  over the cooler region and assuming steady state gives
$$
\int_0 (E+P) \vec{v} \cdot \vec{dS} = -(1-g)n_0^2 \Lambda_0 V_0.
$$
where $V_0$ is the volume of the cooler region.  The integral on the left hand side can be estimated to be $\gamma p (v_{L,{\rm hot}} - v_{L,{\rm cool}}) S_0 /(\gamma-1) \sim (1-f^{-1/3})  (L/L_0)  U/t_{\rm mix, hot}$ ($f^{-1/3}$ factor appears because the velocity in the cooler region is smaller by this factor), where $S_0$ is the surface area of the cooler region and $L_0 \sim V_0/S_0$ is the length-scale of the cooler region, and $t_{\rm mix, hot}= L/v_{L,{\rm hot}}$. The $(L/L_0)$ factor should be replaced by 1 if the driving scale is larger than $L_0$.
Assuming the driving scale to be similar to the core size ($L\approx L_0$), the heating rate due to turbulent mixing is $\sim n_0^2 \Lambda_0 (1-f^{-1/3}) (fg)^{1/3} (c_{s0}t_{{\rm cool},0}/L)^{2/3} \sim n_0^2 \Lambda_0 (1-f^{-1/3})  (c_{s0}t_{{\rm cool},0}/L)^{2/3}$, which can be comparable to the cooling rate for subsonic cooling ($c_{s0}t_{{\rm cool},0} \gtrsim L$) in the cooler region (mimicking the core). Here we have assumed that $(fg)^{1/3} \approx 1$; this holds not only for our choice of $f$ and $f_v$ but also for a wide range of reasonable values. Thermal conduction will also transport heat from hotter to cooler regions without turbulence. The Mach number in the hotter and cooler regions are $\sim (fg)^{1/3}f^{-1/2} (L/c_{s0} t_{{\rm cool},0})^{1/3}$ and  $\sim g^{1/3} (L/c_{s0} t_{{\rm cool},0})^{1/3}$, respectively. For subsonic cooling both zones can have Mach number ${\cal M} \lesssim 1$. The cooler zone is thermally unstable (and cooling on an average at the beginning) and part of it cools to thermally stable temperature and the rest is mixed in the hot phase. This is clearly seen by comparing early and late time pdfs in Figure \ref{fig:pdf}. Most importantly, the Mach number in the hot phase for turbulent mixing runs is small and consistent with the observational limits.

Figure \ref{fig:pdf_mix} shows the volume and mass pdfs as a function of the Mach number for the hydro, MHD and the MHD mixing runs at early and later times. At late times the Mach number pdf for the mixing run is peaked at significantly smaller Mach numbers as compared to the hydro runs. The lower Mach number peak, corresponding to the hot phase, is roughly consistent with the velocity constraints in cool core clusters. 
The higher Mach number peak corresponds to the gas at thermally stable temperature, and cold filaments can indeed have slightly supersonic velocities. Thus, turbulent mixing of hot and cooler ICM via AGN jets is a viable source for heating cool cluster cores.

\subsection{Density dependence of microscopic heating}

Some of the previous work (\citet{sha10,mcc12,sha12}) has added heat in cool core clusters as thermal energy. However, observations of jets expanding in the ICM suggest that heating should be via injection of kinetic energy due to shocks and turbulence. In this paper we have explored the implications of turbulent heating of the ICM. The thermal instability timescale ($t_{\rm TI}$) is not directly measurable from observations  (although $t_{\rm cool}$ is) because it depends on the density dependence of the unknown heating function. The internal energy equation is
$$
\rho T \frac{ds}{dt}  = -n_en_i \Lambda(T) + q^+(n,\vec{r},t),
$$
where $s \equiv k_B \ln(P/\rho^\gamma) /[(\gamma-1)\mu m_p]$ is the specific entropy. The isobaric thermal instability timescale for an above form of the heating function is related to the cooling time via
\be
 \label{eq:TI}
t_{\rm TI} = \frac{\gamma t_{\rm{cool}}}{2-\frac{\rm{d} \ln \Lambda}{\rm{d} \ln \rm{T}}-\alpha},
\ee
where $q^+ \propto n^\alpha$. Thus, $t_{\rm TI} \approx (10/9) t_{\rm cool}$ for $\alpha=0$ and $t_{\rm TI} \approx (10/3) t_{\rm cool}$ for $\alpha=1$ in the free-free regime ($\Lambda \propto T^{1/2}$; see Eq. 19 in \citealt{mcc12} for details). We can thus measure the density dependence of the heating rate by measuring the thermal instability growth rate from numerical simulations.

Simulations of cool core clusters in thermal balance (\citealt{sha12}) show that cold gas can condense out of the hot phase only if $t_{\rm TI}/t_{\rm ff} \lesssim 10$. Moreover, observations (see Fig. 11 in \citealt{mcc12}) show that clusters with $t_{\rm cool}/t_{\rm ff} \lesssim 10$ show evidence for extended cold gas filaments.\footnote{The observed critical value of $t_{\rm cool}/t_{\rm ff}$ may actually be close to 20 rather than 10 because \citet{mcc12} interpreted electron pressure as the total pressure in fitting ICM profiles (private communication with M. McCourt). The existence of a critical value is more important than its precise value.} Thus, if thermal instability is responsible for the observed cold gas filaments in clusters then a comparison of observations and AGN jet simulations with cooling can constrain the microscopic heating mechanism.
In particular, if $t_{\rm TI} \approx t_{\rm cool}$ then $\alpha \approx 0$ and the heating rate per unit volume of the core is roughly constant.

We have calculated the thermal instability timescale from our simulations by measuring the growth rate of the rms density perturbations in the linear thermal instability phase (Fig. \ref{fig:rho}). The measured thermal instability timescales and the corresponding $\alpha$ (c.f. Eq. \ref{eq:TI}) are listed in Table \ref{tab:tab1}. The measured growth for most of our runs are consistent with $\alpha\approx 0$ and a constant heating rate per unit volume. However, there is some variation around this value.

We can make a naive estimate of the density dependence of the heating per unit volume. If turbulent mixing in a medium with background temperature gradient behaves like thermal conduction, then
we do not expect growth for modes at scales smaller than the turbulent Field length. However, we do not expect turbulent mixing to affect the thermal instability growth rate at larger scales. 
Thus, turbulent mixing
is expected to have a similar dependence of heating rate on density as heating due to thermal conduction (see Eq. 8 in \citealt{sha10}); namely, $\alpha \approx 0$.
As already mentioned, this is a crude estimate as the process of turbulent mixing/heating is highly nonlinear, and this quantity should be calculated from numerical simulations.  
Table \ref{tab:tab1} shows that $\alpha \approx 0$ for most of our runs (irrespective of magnetic fields and conduction), a value supported by comparing idealized simulations 
and cluster observations. 

\subsection{How do filaments form?}
All our simulations, whenever they show cold gas, show it in form of clouds and not in form of filaments (see Figs. \ref{fig:snap}, \ref{fig:snap_MHD}) but observations of cold gas in cluster cores show filamentary gas (e.g., \citealt{mcd10}). The question is what are we missing in our simulations that produces cold filaments. We can think of two effects: first, our simulations do not include gravity which makes extended cold gas short-lived (being heavier than its surroundings, cold gas falls toward the center on a free-fall timescale) and filamentary because of ram pressure faced by cold gas falling through the hot ICM (e.g., see the right panel of Fig. 1 in \citealt{mcc12}); second, non-thermal component such as small-scale magnetic fields and (adiabatic) cosmic rays may be required to prevent the collapse of cold gas along the magnetic field lines (this is investigated in more detail in \citealt{sha10}). Also, unlike in our setup with uniform turbulence, cold gas in reality may be condensing out of relatively undisturbed gas.

 One may naively think that anisotropic thermal conduction can lead to long-lived cold filaments elongated along the local magnetic field direction. In the linear regime anisotropic thermal conduction suppresses the growth of modes along field lines for scales smaller than the Field length,  but nonlinearly the cold blobs collapse because radiative cooling overwhelms conductive heating.
The non-thermal pressure of cosmic rays (or tangled magnetic fields) can prevent the collapse of cold gas along field lines, provided the cosmic ray diffusion coefficient is $\lesssim 10^{29}$ cm$^2$s$^{-1}$ (\citealt{sha10}). Cosmic rays compressed in the cold, dense gas are required in the hadronic scenario for the gamma rays emitted by the Fermi bubble in the Galactic center (\citealt{cro13} and references therein). Numerical simulations are still far from the stage where they can reproduce the observed morphology of cold filaments in the ICM.

In conclusion, our numerical simulations show that the scenario in which turbulent heating balances radiative cooling in cluster cores, in order to have condensation of cold filaments, requires a Mach number of order unity (c.f., Eq. \ref{eq:cond_MP}). This is clearly ruled out from observations. The scenario where cool cores are predominantly heated by mixing of hotter gas with the cooler core due to AGN jets, gives reasonable velocities in the hot gas and are consistent with observations (see Fig. \ref{fig:pdf_mix}). This has been pointed out in past by analytic calculations (e.g., \citealt{den05}). Now that AGN jet simulations have become mature enough to achieve thermal balance in cluster cores, the focus should shift on identifying the mechanism via which AGN jets are able to heat up cluster cores. Our paper is a small step in this direction.

\section*{Acknowledgements}
NB wishes to acknowledge the help provided by KVPY for arranging the visit to IISc.
The numerical simulations were carried out on computer cluster provided by the start-up grant of PS at IISc. This work is partly supported by the DST-India grant no. Sr/S2/HEP-048/2012. We thank Mike McCourt for his help with resolving  a crucial bug in our setup and for his comments on the paper. We are grateful to the anonymous referee for comments that significantly improved the paper.
   

{}

\label{lastpage}

\end{document}